\documentclass[aps,prd,twocolumn,floatfix,nofootinbib,superscriptaddress]{revtex4-1}
\usepackage{dcolumn}

\usepackage{amsmath,amsfonts,extarrows,amssymb,mathrsfs,times,bm}
\usepackage{extarrows}
\usepackage{graphicx}
\usepackage{float}
\usepackage{graphicx,epstopdf}
\usepackage{dcolumn}
\usepackage{exscale}
\usepackage{relsize}
\usepackage{mathtools}
\usepackage{mhchem}
\usepackage{xcolor}
\usepackage[colorlinks=true,breaklinks=true,linkcolor=blue,citecolor=blue,urlcolor=blue]{hyperref}
\newcommand{\nn}{\nonumber}

\begin{document}

\title{Gravitational Lensing Using Werner’s Method in Cartesian-like Coordinates}
\author{Zonghai Li}
\email{lizzds@gznu.edu.cn}
\affiliation{School of Physics and Electronic Science, Guizhou Normal University, Guiyang, Guizhou 550001, China}

\date{\today}
\begin{abstract}
The Gibbons-Werner method, which employs the Gauss-Bonnet theorem and the optical/Jacobi metric to compute deflection angles in the weak-field regime, has become widely popular in recent years. Werner extended this method to stationary spacetimes, where the optical/Jacobi metric takes the form of a Finsler metric of Randers type, by adopting an osculating Riemannian metric. Werner's method is particularly valuable as it provides a concise expression for the deflection angle, holds potential for application to gravitational lensing in Finsler spaces beyond the Randers type, and may stimulate the broader application of Finsler geometry across diverse fields. However, it has not been widely adopted due to the cumbersome calculations involved in the conventional coordinates $(r,\phi)$. This paper aims to reduce the computational burden of Werner's method, thereby making it a more practical and accessible approach. To this end, we introduce Cartesian-like coordinates $(X,Y)$ to construct the osculating Riemannian metric and calculate the deflection angle using the Gauss-Bonnet theorem. We demonstrate the effectiveness of our method by computing the deflection of massive particles in Kerr spacetime, rotating Bardeen (Hayward) regular spacetime, and Teo rotating wormhole spacetime. Future research aims to extend Werner’s method to the study of gravitational lensing in Finsler spaces beyond the Randers type. Although this paper focuses on Randers metrics, it provides a foundation for such an extension.

\end{abstract}
\maketitle


\section{Introduction}

In recent years, the geometric method introduced by Gibbons and Werner~\cite{Gibbons-Werner}, utilizing the Gauss-Bonnet theorem to study gravitational lensing, has gained favor among researchers. This method involves the application of a spatial metric called optical metric, which was initially introduced by Weyl in 1917~\cite{Weyl}. According to Fermat's principle in general relativity, the spatial part of a light-like geodesic in 4D spacetime corresponds to a geodesic in the 3D optical metric space (optical geometry)~\cite{Perlick}.
The Gibbons-Werner method was initially applied to the deflection of light rays in static spacetimes such as Schwarzschild spacetime, where the optical metric is Riemannian. By applying the Gauss-Bonnet theorem to the 2D optical geometry (corresponding to the equatorial plane), the deflection angle is elegantly expressed as the integral of the Gaussian curvature of the optical metric.

In stationary spacetime, the optical metric takes the form of a Randers type Finsler metric. To apply the Gauss-Bonnet theorem from Riemannian geometry to study gravitational deflection, two methods have been developed: Werner's method~\cite{Werner2012} and the Ono-Ishihara-Asada (OIA) method~\cite{OIA2017}. In Werner's method, the space where light propagates is defined by the osculating Riemannian metric associated with the optical Randers metric. The osculating Riemannian metric, introduced by Nazım~\cite{Nazım}, maintains the geodesics of the corresponding Finsler metric as its own geodesics. Consequently, the deflection angle is formally expressed in the same manner as in static spacetime, represented as the integral of the Gaussian curvature. However, computations using Werner's method are very cumbersome, as evident from Refs.
~\cite{
	A day so happy,
	Fog lifted early I worked in the garden,
	Hummingbirds were stopping over honeysuckle flowers,
	There was no thing on earth I wanted to possess,
	I knew no one worth my envying him,
	Whatever evil I had suffered I forgot,
	To think that once I was the same man did not embarrass me,
	In my body I felt no pain,
	When straightening up I saw the blue sea and sails}. 
In the OIA method, the light rays space is defined by the Riemannian part of the Randers metric. This is essentially based on the observation that the motion of a free particle within a Randers space can be analogously described as the motion of a charged particle influenced by a magnetic field in a Riemannian space. In this method, the deflection angle is expressed as integrals involving both Gaussian curvature and geodesic curvature. Compared to the Werner's method, the OIA method is more popular and widely applied due to its ease of computation~\cite{OIA-Teo,OIA-monopole,OIA-Ovgun-a,OIA-OSSa,OIA-OSSb,OIA-CrisnejoGR,OIA-KGW-a,OIA-KGW-b,OIA-KSG,OIA-HJJLM,OIA-PGG,OIA-WMG,OIA-BG,OIA-BBB,OIA-AK,OIA-XZSLD,OIA-GaoLiu}.

However, Werner's method has some good characteristics. Firstly, since it involves only the integral of Gaussian curvature, it provides a concise expression for the deflection angle. Secondly, when studying gravitational lensing in Finsler geometry beyond the Randers type, the OIA method, being specifically designed for Randers geometry, is no longer applicable. However, Werner’s method may still be valid, as the osculating Riemannian approach applies to general Finsler geometries and is not restricted to the Randers type. Finally, Werner's method can inspire the application of Finsler geometry in more diverse fields. Considering these favorable characteristics, it would be foolish to let cumbersome calculations hinder the application of Werner's method. Therefore, alleviating the computational burden associated with Werner's method is important, and that is precisely the objective of this paper.

In Ref.~\cite{Li-Zhou}, it has been observed that calculating the deflection angle of light rays in Kerr-Newman spacetime under harmonic coordinates using Werner's method is straightforward. However, this calculation cannot be generalized because obtaining harmonic coordinates for any a stationary spacetime is not straightforward. Can we find simpler coordinates that make calculations with Werner's method more convenient? This paper will address this question and provide a general framework that is applicable to any stationary spacetime. Since our primary focus is on the deflection problem on the equatorial plane, to simplify the process, we will only consider the coordinate transformation on the 2D space rather than the entire 4D spacetime. In particular, we will introduce Cartesian-like coordinates $(X, Y)$ to implement Werner's method, specifically involving the construction of the osculating Riemannian metric of the Randers metric and utilizing the Gauss-Bonnet theorem to study the deflection angle.

In addition to the general construction, we will also demonstrate in this paper some examples of using Werner's method to compute deflection angles in Cartesian-like coordinates $(X, Y)$. Specifically, we will consider the deflection of massive particles (or timelike particles). The lensing of massive particles is richer than that of light, and its study has corresponding theoretical and observational value, attracting the attention of researchers \cite{
	To a Hostess Saying Good Night,
	James Arlington Wright,
	Shake out the ruffle turn and go,
	Over the trellis blow the kiss,
	The dewdrops sway and tremble,
	Some of the guests will never know,
	Another night to shadow this,
	Some of the birds awake in vines,
	Will never see another face,
	So frail so lovely anyplace,
	Between the birdbath and the bines,
	O dark come never down to you,
	I look away and look away,
	Over the moon the shadows go,
	Over your shoulder nebulae,
	Some of the vast the vacant stars,
	Through eons,
	Will never see your face at all,
	Your frail your lovely eyelids fall,
	Between Andromeda and Mars}. 
Using the Jacobi metric for curved spacetime~\cite{Gibbons2016,Chanda2019}, one can extend the application of the Gibbons-Werner method from massless signals to signals with mass~\cite{massiveGB-CG,massiveGB-CGJ,massiveGB-LiHZ,massiveGB-LiJa,massiveGB-LiA,massiveGB-LiZ,massiveGB-CarvalhoAML,massiveGB-CarvalhoAM}, and further, to signals with both mass and charge \cite {massiveGB-CrisnejoGV,massiveGB-LiJb,massiveGB-LiDJ,massiveGB-LiWJ,massiveGB-LiJc}. Since we've mentioned signal extensions, it's worth noting that the extended research on other aspects using the Gibbons-Werner method. Regarding sources-lens distance, finite-distance effects are considered~\cite{ISOA2016,OIA-OA}. Regarding the lenses, the plasma background~\cite{massiveGB-CG,CGR-plasma,Javed-plasma1,Javed-plasma2}, acoustic black holes~\cite{Qiao&Zhou,Molla&Debnath}, and so on are taken into account. Furthermore, improvements to the Gibbons-Werner method itself are being made to accommodate a wider range of scenarios~\cite{TOA,massiveGB-LiZA,HuangCa,HuangCb,HuangCL,HuangSC}. Naturally, various aspects of research may be interconnected. 

The structure of this paper is outlined as follows. In Sec.\ref{Gou-Fang-Pi}, we give the foundational concepts and tools that will be utilized later, encompassing Finsler geometry, osculating Riemannian metric, and Jacobi (optical) metric. In Sec.\ref{Fang-Gou-Pi}, we begin by reviewing Werner's method. Subsequently, we introduce Cartesian-like coordinates $(X,Y)$ to write Werner's method and derive an expression for calculating the deflection angle. To demonstrate the convenience of this expression in applications, we provide three examples in Sec.\ref{Stopping by Woods on a Snowy Evening}, which involve calculations of the deflection angle of massive particles in Kerr spacetime, rotating Bardeen ( Hayward) regular spacetime, and Teo wormhole spacetime, respectively. The conclusion is presented in Sec.\ref{Conclusion}. Throughout this paper, we use geometric units where $c=G=1$.

\section{Preliminaries}
\label{Gou-Fang-Pi}

\subsection{A Little Finsler Geometry}
 
Let $M$ be a smooth manifold of dimension $n$. $T_xM$ denotes the tangent space at $x\in M$. The tangent bundle of $M$ is 
\begin{align}
	TM:=\bigcup_{x\in M}T_xM=\{(x,y)|x\in M,y=y^i\frac{\partial}{\partial x^i}\in T_xM\},\nn
\end{align}
with local coordinates $(x^i, y^i)$.
A Finsler metric is a non-negative function defined on the tangent bundle, $F: TM \to [0, \infty)$, satisfying the following three properties~\cite{Bao-chern-shen}:

\begin{enumerate}
	\item \textit{Smoothness}: $F$ is $C^{\infty}$ on $TM \setminus \{0\}$, where $\{0\}$ denotes the zero section of $TM$.
	
	\item \textit{Positive homogeneity}: $F(x, \lambda y) = \lambda F(x, y)$ for any $\lambda > 0$.
	
	\item \textit{Strong convexity}: The matrix composed of the fundamental tensor
	\begin{align}
	g_{ij}(x, y) = \frac{1}{2} \frac{\partial^2 F^2}{\partial y^i \partial y^j},
	\end{align}
	is positive-definite.
\end{enumerate}

The pair $(M,F)$ is called Finsler manifold or Finsler space. If the fundamental tensor of the Finsler metric $F$ is independent of $y$, or if $F^2$ can be expressed as a quadratic form in terms of $y$, that is, $F(x,y)=\sqrt{g_{ij}(x)y^iy^j}$, then $F$ is a Riemannian metric. In other words, Riemannian geometry is Finsler geometry with the quadratic restriction. 

The Riemannian metric is symmetric, meaning that $F(x,-y)=F(x,y)$. However, this perfect property may render it ineffective in capturing the asymmetries of the real world—such as the one-way nature of time. In 1941, physicist Randers recognized this limitation and introduced an asymmetric metric in 4D spacetime~\cite{Randers}. Subsequently, it was discovered that the metric introduced by Randers possesses a Finsler structure, termed the Randers metric, with the form~\cite{Cheng-Shen}
\begin{align}
	\label{Jiu-Mozhi}
	F(x, y) =\sqrt{\alpha_{ij}(x)y^iy^j}+\beta_i(x)y^i,
\end{align}
where $\alpha_{ij}$ represents the Riemannian metric and $\beta_i$ is a 1-form on $M$, satisfying $\alpha^{ij}\beta_i\beta_j<1$. The fundamental tensor for the Randers metric can be obtained easily as follows
\begin{align}
	\label{Qiao-feng}
	g_{i j}\left(x, y\right)=&\alpha_{i j}+\beta_{i} \beta_{j}-\frac{\left(\beta_{k} y^{k}\right) \alpha_{i k} \alpha_{j l} y^{k} y^{l}}{\left(\alpha_{k l} y^{k} y^{l}\right)^{3 / 2}}\nn\\
	&+\frac{\left(\alpha_{i j} \beta_{k}+\alpha_{j k} \beta_{i}+\alpha_{k i} \beta_{j}\right) y^{k}}{\left(\alpha_{k l} y^{k} y^{l}\right)^{1 / 2}}.
\end{align}

In local coordinates, the equation of a geodesic in Finsler space is given by
\begin{align}
\ddot{x}^i(\tau)+2 G^i\left(x(\tau), \dot{x}(\tau)\right)=0,
\end{align}
where the dot means differentiation with respect to the parameter $\tau$, and $G^i$ are called the geodesic spray coefficients, given by
\begin{align}
G^i(x,y)=\frac{1}{4} g^{i l}(x,y)\left[2\frac{\partial g_{jl} }{\partial x^k}(x,y)-\frac{\partial g_{jk}}{\partial x^l}(x,y)\right]y^jy^k.\nn
\end{align}
For Riemannian metric, the spray coeﬃcients are
\begin{align}
	G^i(x,y)=\frac{1}{2}\Gamma_{jk}^i(x)y^jy^k,\nn
\end{align}
where $\Gamma_{jk}^i$ are the Christoffel symbols.

\subsection{Osculating Riemannian metric}

Along a nowhere vanishing vector field $V$ on $M$, the fundamental tensor induces a Riemannian metric, given by
\begin{align}
	\label{Osculating}
	\bar{g}_{ij}(x)=&g_{ij}(x,V(x)).
\end{align}
This metric $\bar{g}_{ij}$ is referred to as the $V-$osculating Riemannian metric associated with $F$. 

In particular, for a given geodesic $\gamma$ in $(M,F)$, one can choose a vector field such that on the geodesic it equals the tangent vector field, i.e., $V(\gamma)=\dot x$. Then, $\gamma$ is also a geodesic of the induced osculating Riemannian space $(M,\bar{g})$, as can be seen from the following equation~\cite{Werner2012}
\begin{align}
	\ddot{x}^i+2 G^i\left(x, \dot{x}\right)=\ddot{x}^i+2 \bar{G}^i\left(x, \dot{x}\right)=0,
\end{align}
where $\bar{G}^i$ denotes the spray coeﬃcients of $\bar{g}$.

The osculating Riemannian method, initially introduced by Nazım \cite{Nazım} and further developed by Varga and others~\cite{Varga1,Varga2,Rund}, now finds significant applications in comparison Finsler geometry~\cite{Ohta}. In physics, it is utilized in gravitational theory and cosmology \cite{osculating-Asanov1,osculating-Asanov2,osculating-KSS,osculating-HHS,osculating-BCHHSM}. Werner~\cite{Werner2012} employed the osculating Riemannian metric and the Gauss-Bonnet theorem to investigate gravitational lensing within stationary spacetimes, a topic relevant to this paper and will be discussed in detail later.

\subsection{Jacobi (Optical) metric}
In Boyer-Lindquist coordinates $(t, r, \theta, \phi)$, the metric for the 4D stationary spacetime is given by
\begin{align}
	\label{BL-metric}
	d s^2=&g_{t t}(r,\theta) d t^2+2 g_{t \phi}(r,\theta) d t d \phi+g_{rr}(r,\theta) dr^2\nn\\
	&+g_{\theta\theta}(r,\theta) d\theta^2+g_{\phi\phi}(r,\theta) d\phi^2.
\end{align}
To investigate the propagation and deflection of particles in this background, we employ one of the fundamental tools of geometric dynamics, namely the Jacobi metric. According to the Maupertuis principle, the spatial trajectories of particles moving in 4D spacetime are geodesics in the Jacobi metric space. For a neutral particle with mass $m$ and conserved energy $E$, the corresponding Jacobi metric takes the form of Eq. \eqref{Jiu-Mozhi}, indicating it is a Randers metric, given by~\cite{Chanda2019}
\begin{align}
	\label{Jacobi-metric}
	F(x,dx)=d\rho=\sqrt{\alpha_{ij}dx^idx^j}+\beta_idx^i,
\end{align}
where
\begin{subequations}
	\label{Jacobi-Randers-ab}
	\begin{align}
		\label{Jacobi-Randers-a}
		&\alpha_{ij}=\frac{E^2+m^2{g}_{tt}}{-{g}_{tt}}\left({g}_{ij}-\frac{{g}_{ti}{g}_{tj}}{{g}_{tt}}\delta_{\phi}^i\delta_{\phi}^j\right),\\
		\label{Jacobi-Randers-b}
		&\beta_i=-E\frac{g_{ti}}{g_{tt}}\delta_{\phi}^i.
	\end{align}
\end{subequations}
In the above, the energy $E$ can be expressed in terms of the asymptotic velocity $v$ of the particle as follows
\begin{align}
	\label{LED}
	E=\frac{m}{\sqrt{1-v^2}}.
\end{align}
By setting $m=0$ and $E=1$, the Jacobi metric given by Eqs.~\eqref{Jacobi-metric}-\eqref{Jacobi-Randers-ab} reduces to the optical metric. Geodesics of the optical metric correspond to spatial light rays, making it highly suitable for studying light propagation and gravitational lensing~\cite{Gibbons-Werner,Werner2012}. Since the optical metric is a special case of the Jacobi metric, for the sake of generality, this paper considers massive particles and the Jacobi metric. The results for light rays can be obtained from the results for particles by letting the particle velocity $v\to1$.

\section{Werner's method in Cartesian-like Coordinates}
\label{Fang-Gou-Pi}

\subsection{Werner's method}

Now, let's examine the deflection of particles in stationary spacetime, particularly focusing on the equatorial plane $(\theta=\pi/2)$ and within the weak-field approximation. The particle trajectory is a geodesic in the 2D Jacobi-Randers space $(\mathcal{M}^{2},F)$, with the metric
	\begin{align}
		\label{Ding-Jacobi}
	F(r,\phi,dr,d\phi)=\sqrt{\alpha_{rr}dr^2+\alpha_{\phi\phi}d\phi^2}+\beta_{\phi}d\phi.
	\end{align}
where $\alpha_{rr}$	, $\alpha_{rr}$, and $\beta_{\phi}$ are given by Eq. \eqref{Jacobi-Randers-ab}.

\begin{figure}[htp!]
	\centering
	\includegraphics[width=8.0cm]{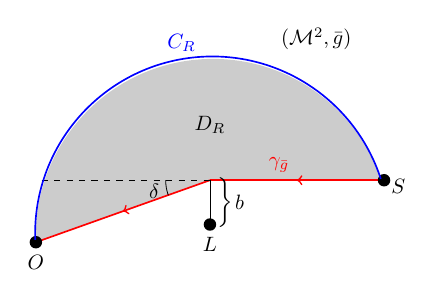}
	\caption{A region $D_{R}\subset (\mathcal{M}^{2},\bar{g})$ with boundary $\partial D_{R}=\gamma_{\bar{g}} \cup C_{R}$. Here, $C_{R}$ denotes a curve defined by $r=R=constant$. Additionally, $\gamma_{\bar{g}}$ represents the trajectory of a particle emitted from the source $S$, deflected by the lens $L$, and arriving at the observer $O$, with a small deflection angle  $\delta$. Moreover, $b$ is the impact parameter.\label{XuWei-01}}
\end{figure}

Werner~\cite{Werner2012} demonstrated that we can employ Nazım's method to construct the osculating Riemannian space $(\mathcal{M}^{2},\bar g)$ by adopting the tangent vector field of the geodesic. As a conclusion, the geodesic in $(\mathcal{M}^{2},F)$ also serves as a geodesic in $(\mathcal{M}^{2},\bar g)$. Therefore, we can investigate particle deflection using the Riemannian space $(\mathcal{M}^2,\bar g)$. In $(\mathcal{M}^{2},\bar{g})$ with coordinates $(r,\phi)$, a particle with trajectory  departs from the source $S$, undergoes deflection by the lens $L$, and arrives at the observer $O$. The trajectory of the particle is denoted as $\gamma_{\bar{g}}$, with a small deflection angle $\delta$, as depicted in Fig. \ref{XuWei-01}. 

Now, we consider a non-singular region $D_R\subset (\mathcal{M}^{2},\bar{g})$ with boundary $\partial D_{R}=\gamma_{\bar g} \cup C_{R}$, where $C_{R}$ is a curve defined by $r=R=constant$, as shown in Fig. \ref{XuWei-01}. Let $\bar{K}$ denote the Gaussian curvature of the Riemannian metric $\bar{g}$, $\chi$ represent the Euler characteristic of $D_R$, and $k$ stand for the geodesic curvature of $\partial D_R$. Applying the Gauss-Bonnet theorem~\cite{Gibbons-Werner} to the region $D_R$, we have
\begin{align}
	\label{GBT}
	\iint_{D_{R}} \bar{K} d S+\int_{S}^{O}{k}(C_R)dl+\varphi_{O}+\varphi_{S}=2\pi,
\end{align}
where $\varphi_{O}$ and $\varphi_{S}$ are the exterior angles of the intersection points of curves $\gamma_{\bar g}$ and $C_R$ at $O$ and $S$ respectively, in the positive sense. In addition, $dS$ represents the area element of $D_R$ and $dl$ denotes the line element along $\partial{D}$. In Eq.~\eqref{GBT}, we utilized $\chi=1$ because $D_R$ is a non-singular region. Additionally, we employed $\oint_{\partial{D}_R}{k}dl=\int_S^O k(C_R)dl$, since $\gamma_{\bar g}$ is a geodesic in $(\mathcal{M}^{2},\bar{g})$ resulting in $k(\gamma_{\bar g})=0$.

Assuming $(\mathcal{M}^{2},\bar g)$ is asymptotically Euclidean space, and the source $S$ and observer $O$ are situated at the asymptotic region. We orient the coordinates such that the angular coordinate of the source is $\phi_S=0$ as $R\to \infty$, and consequently, the angular coordinate of the observer becomes $\phi_O=\pi+\delta$. In this limit, we have $\varphi_O+\varphi_S\to\pi$, and $k(C_R) dl\to d\phi$. Taking these facts into account, letting $R\to\infty$, Eq.~\eqref{GBT} leads to
\begin{align}
	\label{shalihuayuan}
	&\iint_{D_{\infty}}  \bar{K} d S+ \int_0^{\pi+\delta}d\phi+\pi=2\pi,
\end{align}
or
\begin{align}
	\label{Gibbons-Werner}
	\delta=&-\iint_{D_{\infty}} \bar{K}dS.
\end{align}
This equation demonstrates that the deflection angle remains independent of the choice of coordinates. 

The asymptotic deflection angle \eqref{Gibbons-Werner} can be further expanded in the coordinates $(r,\phi)$ as
\begin{align}
	\label{DAF}
	\delta^{[n]}=&-\int_{0}^{\pi+\delta^{[n-2]}}\int_{\{r(\phi)\}^{[n-1]}}^\infty\left\{ \bar{K}\sqrt{|\bar{g}|}\right\}^{[n]}~drd\phi,
\end{align}
where $r(\phi)$ represents the particle trajectories and $|\bar{g}|$ denotes the determinant of $\bar{g}$. Here, we use the superscript $[i]$ on a quantity to indicate that it is accurate up to order $i$. Note that in the above formula, when $ n = 1 $, the term $\delta^{[n-2]}$ results in a negative order. However, in this case, lower-order deflection angle information is not actually needed. To address this, we define $[i] = [0]$ for $ i < 0 $.
 Thus, to compute the deflection angle up to order $n$ in coordinates $(r,\phi)$, we require not only information about the $(n-1)$th-order particle trajectory but also information about the $(n-2)$th-order deflection angle~\cite{massiveGB-CGJ}. This process involves iteration; for example, to compute the third-order deflection angle, it is necessary to first calculate the first-order deflection angle \cite{massiveGB-LiWJ,massiveGB-LiJc}.

\subsection{Osculating Riemannian metric in coordinates $(X,Y)$}

This subsection will discuss the construction of the Riemannian metric in Cartesian-like coordinates, and in the following subsection, we will provide expressions for the deflection angle in this coordinates. Let's introduce the Cartesian-like coordinates $(X,Y)$ as follows
\begin{align}
	\label{PC}
	&X=r\cos\phi,\quad Y=r\sin\phi.
\end{align}
Then, we have
\begin{subequations}
	\label{PCT}
	\begin{align}
		&dr=\frac{1}{r}\left(XdX+YdY\right),\\
		&d\phi=\frac{1}{r^2}\left(-YdX+XdY\right),
	\end{align}
\end{subequations}
where $r=\sqrt{X^2+Y^2}$.
Using Eqs.~\eqref{PC} and \eqref{PCT}, we can express the Jacobi-Randers metric \eqref{Ding-Jacobi} as
\begin{align}
	F(X,Y,dX,dY)=&\sqrt{\alpha_{IJ}dx^Idx^J}+\beta_{I}d{x^I},
\end{align}
where $x^I\in\{X,Y\}$ and the metric components are
\begin{subequations}
	\label{Randersxyz}
	\begin{align}
	&	\alpha_{XX}
		=\frac{1}{r^4}(\alpha_{rr}X^2r^2+\alpha_{\phi\phi}Y^2),\\
	&	\alpha_{XY}
		=\frac{1}{r^4}(\alpha_{rr}r^2-\alpha_{\phi\phi})XY,\\
	&	\alpha_{YY}
		=\frac{1}{r^4}(\alpha_{rr}Y^2r^2+\alpha_{\phi\phi}X^2),\\
	&	\beta_X=-\frac{\beta_\phi Y}{r^2},\quad \beta_Y=\frac{\beta_\phi X}{r^2}.
	\end{align}
\end{subequations}

To construct the osculating Riemannian metric, we employ the straight line approximation, which corresponds to the zeroth-order particle trajectory, in coordinates $(X,Y)$, represented by $Y(X)=b$, with $b$ being the impact parameter, as depicted in Fig. \ref{XuWei-02}. With this approximation, we can choose the following vector filed
\begin{align}
	\label{Yong_chun-zhang_tian_zhi}
V^X=-1,\quad V^Y=0.
\end{align}

Substituting $(\alpha_{IJ},\beta_I)$ into Eq. \eqref{Qiao-feng} yields the fundamental tensor $g_{IJ}(X^I,V^I)$. Then, by employing the vector field given in Eq. \eqref{Yong_chun-zhang_tian_zhi} , we obtain the osculating Riemannian metric $\bar g_{IJ}(X^I)=g_{IJ}(X^I,V^I(X^I))$ as follows
\begin{align}
	\label{The woods are lovely, dark and deep}
	\bar{g}_{IJ}=&\alpha_{IJ}+\beta_{I} \beta_{J}-\frac{\alpha_{X I} \beta_{J}+\alpha_{J X} \beta_{I}}{\sqrt{\alpha_{XX}}}\nn\\
		&+\frac{\left(\alpha_{I X} \alpha_{J X}-\alpha_{IJ}\alpha_{XX}\right)\beta_{X}}{(\alpha_{XX})^{3 / 2}},
\end{align}
Written in detail, it is as follows
\begin{subequations}
	\label{Two roads diverged in a yellow wood}
\begin{align}
	\bar{g}_{XX}=&\left(\sqrt{\alpha_{XX}}-\beta_{X}\right)^2,\\
	\bar{g}_{XY}=&\left(\alpha_{XY}-\sqrt{\alpha_{XX}}\beta_{Y}\right)\left(1-\frac{ \beta_{X}}{\sqrt{\alpha_{XX}}}\right),\\
	\bar{g}_{YY}=&\alpha_{YY}+\left(\beta_{Y}-\frac{2\alpha_{X Y} }{\sqrt{\alpha_{XX}}}\right)\beta_{Y}\nn\\
	&+\frac{\left[(\alpha_{XY})^2-\alpha_{XX}\alpha_{YY} \right]\beta_{X}}{(\alpha_{XX})^{3 / 2}},
\end{align}
\end{subequations}
where $\alpha_{IJ}(X,Y)$ and $\beta_I(X,Y)$ are given by Eq. \eqref{Randersxyz}.
	
\subsection{Deflection angle}
\begin{figure}[t]
	\centering
	\includegraphics[width=8.5cm]{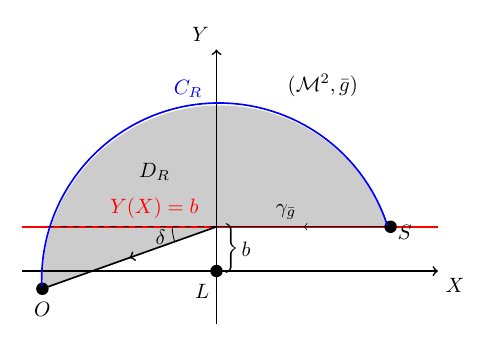}
	\caption{The region $D_R \subset (\mathcal{M}^2, \bar{g})$ with coordinates $(X,Y)$. The straight line approximation is given by $Y(X)=b$. \label{XuWei-02}}
\end{figure}

The deflection angle \eqref{Gibbons-Werner} is invariant under coordinate transformations. Now, let's consider its expression in the coordinates $(X,Y)$. Our goal is to calculate the leading-order deflection angle, thus we only need to use the straight line approximation $Y(X)=b$ (We have already utilized it once when constructing the osculating Riemannian metric), as illustrated in Fig.~\ref{XuWei-02}. Note that in this approximation, as $R\to\infty$, we have $D_\infty=\{(X,Y) \in \mathcal{M}^2 \mid b \leq Y < \infty\}$. Thus, the deflection angle \eqref{Gibbons-Werner} can be written as
	\begin{align}
		\label{DA-XY}
		\delta= & -\int_{-\infty}^{\infty}\int_{b}^{\infty}\bar{K}\sqrt{|\bar{g}|}~dYdX.
	\end{align}
The Gaussian curvature can be calculated by
\begin{align}
	\label{Gauss}
	\bar{K}=\frac{\bar{R}_{XYXY}}{|\bar g|},
\end{align}
where $\bar{R}_{IJKL}$ is the purely covariant Riemannian curvature tensor associated with $\bar g$. 

Regarding Eq.~\eqref{DA-XY}, several points should be noted. First, similar to Eq.~\eqref{Gibbons-Werner}, it is only applicable to asymptotically Euclidean spaces. Second, while Eq.~\eqref{Gibbons-Werner} provides a complete expression for the deflection angle, Eq.~\eqref{DA-XY} is an approximation valid only at the leading order. Lastly, Eq. \eqref{DA-XY} may cause some confusion regarding Fig. \ref{XuWei-02}, as it omits the sector region containing $\delta$. However, at the leading order, the deflection information is fully encoded in the Gaussian curvature $\bar{K}$, which distinguishes it from a flat space.

For asymptotically non-Euclidean spaces, terms like $k(C_R)$ in Eq. \eqref{GBT} need adjustments~\cite{A day so happy, I knew no one worth my envying him}, thus requiring improvements to the deflection angle \eqref{Gibbons-Werner}. The refined deflection angle can be easily transformed into the coordinates $(X,Y)$. To compute higher-order deflection angles, it's necessary to consider curved particle trajectories beyond the straight line approximation. This improves the osculating Riemannian metric in Eq. \eqref{The woods are lovely, dark and deep} and the integration domain in deflection angle \eqref{DA-XY}. Specifically, the deflection angle up to order $n$ can be expressed as
\begin{align}
	\label{KangjuHLbyT}
	\delta^{[n]}= & -\int_{-\infty}^{\infty}\int_{\left\{Y(X)\right\}^{[n-1]}}^{\infty}\left\{\bar{K}\sqrt{|\bar{g}|}\right\}^{[n]}~dYdX.
\end{align}
Compared with expression \eqref{DAF}, we can observe that computing higher-order deflection angles in coordinates $(X,Y)$ is more convenient than in the polar coordinates $(r,\phi)$. In the coordinates $(X,Y)$, we only require the low-order particles trajectories, without the need to iterate low-order deflection angles. This convenience in computing high-order deflection angles applies not only to Werner's method but also to all methods utilizing the Gauss-Bonnet theorem.

Finally, why don't we directly transform an osculating Riemannian metric already constructed in the coordinates $(r,\phi)$ into the coordinates $(X,Y)$, instead of starting from the Jacobi-Randers metric? The form of the geodesic tangent vector field~\eqref{Yong_chun-zhang_tian_zhi} in the coordinates $(r,\phi)$ is more complex, which makes it difficult to construct the osculating Riemannian metric. If we want to compute higher-order deflection angles, we must choose a construction beyond the straight line approximation, which will lead to more difficulties. Even if we construct the osculating Riemannian metric, transforming it into the coordinates $(X,Y)$ is not easy due to its complex form in the coordinates $(r,\phi)$. Therefore, we begin with the Jacobi-Randers metric and then introduce the coordinates $(X,Y)$ to describe Werner's method.

\section{Three Examples}
\label{Stopping by Woods on a Snowy Evening}
This section provides several examples of calculating the deflection angle of massive particles using Eq.~\eqref{DA-XY}. We will examine Kerr spacetime, Bardeen (Hayward) rotating spacetime, and Teo wormhole spacetime. These spacetimes are selected not only for their typicality but also because previous works have applied Werner's method to compute the deflection angle in the coordinates $(r, \phi)$ for these spacetimes. This allows us to compare both the computational process and the results with these works.

\begingroup
\renewcommand{\thesubsection}{Example \arabic{subsection}}
\subsection{Kerr spacetime}

The Kerr metric describes the spacetime outside a rotating body with mass $M$ and angular momentum per unit mass $a$. In Boyer-Lindquist coordinates $(t, r, \theta, \phi)$, its line element is given by~\cite{Kerr-BL}

\begin{align}
	\label{Whose woods these are I think I know}
	ds^2=&-\left(1-\frac{2Mr}{\Sigma}\right)dt^2+\frac{\Sigma}{\Delta}dr^2+\Sigma d\theta^2\nn\\
	&+\frac{1}{\Sigma}\left[\left(r^2+a^2\right)^2-\Delta a^2\sin^2\theta\right]\sin^2\theta d\phi^2\nn\\
	&-\frac{4Mar}{\Sigma}\sin^2\theta dt d\phi,
\end{align}

where
\begin{align}
	\Sigma=r^2+a^2\cos^2\theta,\quad\Delta=r^2-2Mr+a^2.\nn
\end{align}

Substituting the metric components of the Kerr spacetime into Eqs.~\eqref{Jacobi-Randers-a} and \eqref{Jacobi-Randers-b}, and considering the equatorial plane $(\theta=\pi/2,~d\theta=0)$, we obtain the following 2D Kerr-Jacobi-Randers metric
\begin{subequations}
	\label{His house is in the village though}
{\small	\begin{align}
		&\alpha_{ij}dx^idx^j=\left(\frac{E^2r^2}{\Delta-a^2}-m^{2}\right)\left[\frac{r^2}{\Delta}dr^2+\frac{r^2\Delta d\phi^{2}}{\Delta-a^2}\right],\\
		&\beta_{i}dx^i=-\frac{2E Ma r}{\Delta-a^2}d\phi.
	\end{align}}
\end{subequations}
	
Applying Eq.~\eqref{Randersxyz}, we can express the Kerr-Jacobi-Randers metric in the coordinates $(X, Y)$. However, this is just a minor intermediate step, and it is unnecessary to write out the result. What's important is to use it according to Eq. \eqref{The woods are lovely, dark and deep} or \eqref{Two roads diverged in a yellow wood} to obtain the osculating Riemannian metric. The final result is as follows
	\begin{subequations}
		\label{He will not see me stopping here}
		\begin{align}
			\bar{g}_{XX}=&E^2v^2\left[1+\frac{2 M}{r}\left(\frac{1}{v^2}+ \frac{X^2}{r^2}\right)\right]\nn\\
			&-\frac{4 M a E^2vY}{r^3}+\mathcal{O}\left(M^2,a^2\right), \\
			\bar{g}_{XY} =&\frac{2 E^2v^2 M X}{r^3}\left(Y+\frac{a}{v}\right)+\mathcal{O}\left(M^2,a^2\right),  \\
			\bar{g}_{YY} =&E^2v^2\left[1+\frac{2 M}{r}\left(\frac{1}{v^2}+ \frac{Y^2}{r^2}\right)\right]\nn\\
			&-\frac{2MaE^2vY}{r^3}+\mathcal{O}\left(M^2,a^2\right). 
		\end{align}
	\end{subequations}
The osculating Riemannian metric is asymptotic Euclidean because $\bar{g}\to E^2v^2(dX^2+dY^2)$ as $r\to \infty$. Therefore, we can use Eq.~\eqref{DA-XY} to calculate the deflection angle.

The determinant of the Riemannian metric~\eqref{He will not see me stopping here} is
{\small
\begin{align}
	\label{To watch his woods fill up with snow}
	|\bar{g}|=&E^4v^4\left[1+2\left(1+\frac{2}{v^2}-\frac{3aY}{vr^2}\right)\frac{M}{r}\right]+\mathcal{O}\left(M^2,a^2\right).
\end{align}}
The Gaussian curvature can be calculated using Eq. \eqref{Gauss}, resulting in
{\small
\begin{align}
\label{And miles to go before I sleep}
\bar{K}=&-\frac{M}{E^2 r^3 v^2}\left[1+\frac{1}{v^2}-\frac{3 a Y}{v r^5}\left(1+\frac{5 X^2}{r^2}\right)\right]+\mathcal{O}\left(M^2,a^2\right).
\end{align}
}
Substituting Eqs. \eqref{To watch his woods fill up with snow} and \eqref{And miles to go before I sleep} into Eq. \eqref{DA-XY}, the calculation simplifies significantly, as shown below:
{\small
	\begin{align}
		\delta=&-\int_{-\infty}^{\infty}\int_{b}^{\infty}\bar{K}\sqrt{|\bar{g}|}~dYdX\nn\\
		=&\int_{-\infty}^{\infty}\int_{b}^{\infty}\frac{M}{r^3}\left[1+\frac{1}{v^2}-\frac{3 a Y}{v r^5}\left(1+\frac{5 X^2}{r^2}\right)\right]dYdX\nn\\
		=&2\left(1+\frac{1}{v^2}\right)\frac{M}{b}-\frac{4Ma}{b^2v}+\mathcal{O}\left(M^2,a^2\right).
		\end{align}
	}
The result corresponds to prograde particle trajectories. Due to the asymmetry of the Randers metric, i.e., $F(x,dx) \neq F(x,-dx)$, the retrograde trajectory is no longer a geodesic of $ F(x,dx) $ given in Eq. \eqref{His house is in the village though}, but instead follows the geodesic of its inverse metric $ F(x,-dx) $. However, there is no need to perform another calculation using the inverse metric. A simple observation reveals that $F(x,-dx,a) = F(x,dx,-a)$. Thus, the retrograde deflection angle can be obtained simply by replacing $ a $ with $-a $. For the connection between the asymmetry of Finsler geometry and the deflection of prograde and retrograde particles, we refer readers to Refs. \cite{massiveGB-LiWJ, massiveGB-LiJc}. By considering both prograde and retrograde cases, the deflection angle can be expressed as
\begin{align}
	\label{DA-Kerr}
	\delta=2\left(1+\frac{1}{v^2}\right)\frac{M}{b}\pm\frac{4Ma}{b^2v}+\mathcal{O}\left(M^2,a^2\right),
\end{align}
where the positive and negative signs are for retrograde and prograde particle trajectories, respectively. 

The result \eqref{DA-Kerr} has been discovered using different methods in various references~\cite{The dewdrops sway and tremble,Some of the guests will never know,Will never see another face,Through eons,massiveGB-CGJ,massiveGB-LiJa}. In particular, Ref. \cite{massiveGB-CGJ} employed Werner's method with coordinates $(r,\phi)$. In comparison, the forms of the osculating Riemannian metric and its Gaussian curvature in the coordinates $(X,Y)$ presented in this paper are more concise, and the integration process is also easier.

Letting $v\to1$, \eqref{DA-Kerr} yields the light deflection angle given by
\begin{align}
	\delta=\frac{4M}{b}\pm\frac{4Ma}{b^2}+\mathcal{O}\left(M^2,a^2\right).
\end{align}
In Ref. \cite{Werner2012}, Werner derived this result using his method in the coordinates $(r,\phi)$. 

\subsection{Rotating Bardeen (Hayward) regular black hole}
The metric of the rotating Bardeen regular black hole spacetime is given by~\cite{regularBH}
{\small
	\begin{align}
		\label{Bardeen}
		d s^2=&-\left(1-\frac{2 M_B r}{\Sigma}\right) d t^2-\frac{4 a M_B r \sin ^2 \theta}{\Sigma} d t d \phi\nn\\
		&+\frac{\Sigma}{\Delta} d r^2+\Sigma d \theta^2+\left(r^2+a^2+\frac{2 a^2 M_B r \sin ^2 \theta}{\Sigma}\right) d \phi^2,
	\end{align}
}
where
\begin{subequations}
	\label{subBardeen}
\begin{align}
	\Sigma & =r^2+a^2 \cos ^2 \theta, \quad \Delta =r^2-2 M_B r+a^2, \\
	M_B & =M\left(\frac{r^2}{r^2+Q_{B}^2}\right)^{3 / 2},
\end{align}
\end{subequations}
and $Q_{B}$ represents the magnetic charge arising from the non-linear electromagnetic field. If we replace $M_B$ with $M_H$ in metrics \eqref{Bardeen}-\eqref{subBardeen}, we obtain the rotating Hayward spacetime~\cite{regularBH}, where
\begin{align}
 M_H = M\frac{r^r}{r^3+Q_H^3},
\end{align}
with $Q_H$ being the deviation parameter. Below, we present the details of calculating the deflection angle in the rotating Bardeen spacetime, while for the rotating Hayward spacetime, we simply provide the result.

The form of the Bardeen-Jacobi-Randers metric is the same as Eq. \eqref{His house is in the village though}, with the substitution of $M$ by $M_B$. More, the osculating Riemannian metric can also be obtained from Eq. \eqref{He will not see me stopping here} by making the same substitution, yielding
	\begin{subequations}
	\begin{align}
		\bar{g}_{XX}=&E^2v^2\left[1+\frac{2 M_B}{r}\left(\frac{1}{v^2}+ \frac{X^2}{r^2}\right)\right]\nn\\
		&-\frac{4 M_B a E^2vY}{r^3}+\mathcal{O}\left(M^2,a^2\right), \\
		\bar{g}_{XY} =&\frac{2 E^2v^2 M_B X}{r^3}\left(Y+\frac{a}{v}\right)+\mathcal{O}\left(M^2,a^2\right),  \\
		\bar{g}_{YY} =&E^2v^2\left[1+\frac{2 M_B}{r}\left(\frac{1}{v^2}+ \frac{Y^2}{r^2}\right)\right]\nn\\
		&-\frac{2M_BaE^2vY}{r^3}+\mathcal{O}\left(M^2,a^2\right). 
	\end{align}
\end{subequations}
It is easy to see that this osculating Riemannian metric is asymptotically Euclidean. Its determinant can also be obtained by a simple substitution from Eq. \eqref{To watch his woods fill up with snow}, yielding

\begin{align}
	\label{My little horse must think it queer}
	|\bar{g}|=&E^4v^4\left[1+2\left(1+\frac{2}{v^2}-\frac{3aY}{vr^2}\right)\frac{M_B}{r}\right]+\mathcal{O}\left(M^2,a^2\right).
\end{align}

However, the Gaussian curvature cannot be obtained from the results of the Kerr spacetime through simple substitution. The result is as follows
\begin{align}
	\label{To stop without a farmhouse near}
\bar{K}=&-\frac{M_B}{E^2 v^2 r^3\left(r^2+Q_B^2\right)^2}\bigg[\left(1+\frac{1}{v^2}\right) r^4\nn\\
&-\frac{3 a Y}{v}\left(6 X^2+Y^2-9 Q_B^2\right)-\left(1+\frac{10}{v^2}\right) Q_B^2 r^2\nn\\
&-2 Q_B^4\left(1-\frac{2}{v^2}\right)\bigg]+\mathcal{O}\left(M^2,a^2\right).
\end{align}

Substituting the determinant given by Eq. \eqref{My little horse must think it queer} and the Gaussian curvature given by Eq. \eqref{To stop without a farmhouse near} into Eq. \eqref{DA-XY}, the deflection angle can be easily calculated, resulting in
	\begin{align}
		\label{DA-Bardeen}
		\delta=&\frac{2 M}{\left(b^2+Q_B^2\right)^2}\left[b\left(b^2+Q_B^2+\frac{b^2-Q_B^2}{v^2}\right)\right.\nn\\
		&\left.\pm\frac{2 a\left(b^2-Q_B^2\right)}{v}\right]+\mathcal{O}\left(M^2,a^2\right),
	\end{align}
where the positive and negative signs are for retrograde and prograde particle trajectories, respectively. Setting $v=1$, the above equation yields the result for light deflection
\begin{align}
	\label{DA-BardeenNull}
	\delta=&\frac{4 M}{\left(b^2+Q_B^2\right)^2}\left[b^3\pm a\left(b^2-Q_B^2\right)\right]+\mathcal{O}\left(M^2,a^2\right)\nn\\
	=&\frac{4M}{b}-\frac{8MQ_B^2}{b^3}\pm\frac{4Ma}{b^2}+\mathcal{O}\left(\frac{1}{b^4}\right).
\end{align}

Using the same calculation manner as for the rotating Bardeen regular spacetime, we can obtain the deflection angle of massive particles in the rotating Hayward spacetime, as follows
\begin{align}
	\label{DA-Hayward}
	\delta=&2\left(1+\frac{1}{v^2}\right)\frac{M}{b}-3\pi\left(\frac{1}{8}+\frac{1}{2 v^2}\right) \frac{MQ_H^3}{b^4}\nn\\
	&\pm\frac{Ma}{b^2v}\left(4-\frac{3Q_H^3\pi}{b^3}\right)+\mathcal{O}\left(M^2,a^2,Q_H^4\right),
\end{align}
and for light rays
\begin{align}
	\label{DA-HaywardNull}
	\delta=&\frac{4M}{b}-\frac{15\pi MQ_H^3}{8b^4}\pm\frac{Ma}{b^2}\left(4-\frac{3Q_H^3\pi}{b^3}\right)\nn\\
	&+\mathcal{O}\left(M^2,a^2,Q_H^4\right).
\end{align}

The results in Eq. \eqref{DA-BardeenNull} and Eq. \eqref{DA-HaywardNull} respectively align with Eq. (3.14) and Eq. (4.15) derived by Werner's method with coordinates $(r,\phi)$ in Ref.~\cite{Hummingbirds were stopping over honeysuckle flowers}. Although only the gravitational deflection of massless particles is considered in Ref.~\cite{Hummingbirds were stopping over honeysuckle flowers}, its calculations are even more cumbersome than those we perform here for massive particles.
 
\subsection{Rotating Teo wormhole}

The Teo metric describes the stationary, axisymmetric traversable wormhole spacetime, which is the rotating generalization of the static Morris-Thorne wormhole, as given by~\cite{Teo}
\begin{align}
	\label{Teo-a}
	ds^2=&-\mathcal{N}^2 dt^2+\frac{dr^2}{1-\frac{\mathcal{B}}{r}}+r^2 \mathcal{H}^2\left[d\theta^2+\sin^2 \theta(d\phi-wdt)^2\right],
\end{align}
where $\mathcal{N}$, $\mathcal{B}$, $\mathcal{H}$ and $w$ are functions of $r$ and $\theta$. In particular, we choose~\cite{Teo,Fog lifted early I worked in the garden,OIA-Teo,massiveGB-LiJa}
\begin{align}
	\label{Teo-b}
	&\mathcal{B}(r)=b_0,\quad \mathcal{N}=\mathcal{H}=1+\frac{(4J \cos\theta)^2}{b_0^3~r},\quad \omega=\frac{2J}{r^3}.
\end{align}
Here, $J$ represents the total angular momentum of the wormhole, and $b_0$ denotes the throat radius of the wormhole with $b_0\leq r$. The spacetime describes two identical, asymptotically flat regions joined together at the throat, $r = b_0$.

Substituting the Teo metric \eqref{Teo-a} with \eqref{Teo-b} into Eqs. \eqref{Jacobi-Randers-a}-\eqref{Jacobi-Randers-b}, and focusing on the equatorial plane $(\theta=\pi/2)$, we can derive the Teo-Jacobi-Randers metric as follows
\begin{subequations}
\begin{align}
&	\alpha_{ij}dx^i dx^j=\left( \frac{E^2}{ 1-r^2\omega ^2}-m^2 \right)\left[ \frac{{dr}^2}{1-\frac{b_0}{r}}+\frac{r^2d\phi^2}{1-r^2w^2}\right],\\
&{\beta_idx^i=-\frac{Er^2\omega  \,\,d\phi}{1-r^2\omega ^2}}.
\end{align}
\end{subequations}
By employing Eq.~\eqref{Randersxyz}, we can express the Teo-Jacobi-Randers metric described above in terms of the coordinates $(X, Y)$. Then, by utilizing either Eq. \eqref{The woods are lovely, dark and deep} or \eqref{Two roads diverged in a yellow wood}, we can derive the corresponding osculating Riemannian metric as follows
\begin{subequations}
		\label{Between the woods and frozen lake}
	\begin{align}
			\bar{g}_{XX}=&E^2 v^2\left(1+b_0 \frac{X^2}{r^3}-\frac{4 J Y}{r^3 v}\right)+\mathcal{O}\left(\epsilon^2\right), \\
			\bar{g}_{XY}=&E^2 v^2\frac{X}{r^3}\left(b_0 Y+\frac{2 J}{v}\right)+\mathcal{O}\left(\epsilon^2\right),  \\
			\bar{g}_{YY}=&E^2 v^2\left(1+b_0 \frac{Y^2}{r^3}-\frac{2 J Y}{r^3 v}\right)+\mathcal{O}\left(\epsilon^2\right),
	\end{align}
\end{subequations}
where $\epsilon^2\in \{b_0^2,b_0J,J^2\}$. Obviously, the Riemannian metric is asymptotically Euclidean.

The determinant of the Riemannian metric \eqref{Between the woods and frozen lake} is given by
\begin{align}
	\label{The darkest evening of the year}
	|\bar g|=E^4v^4\left(1+\frac{b_0}{r}-\frac{6JY}{r^3v}\right)+\mathcal{O}\left(\epsilon^2\right),
\end{align}
and the Gaussian curvature is expressed as
\begin{align}
	\label{He gives his harness bells a shake}
\bar{K}=-\frac{1}{2E^2 r^3 v^2}\left[b_0-\frac{6JY}{vr^2}\left(1+\frac{5 X^2}{r^2}\right)\right]+\mathcal{O}\left(\epsilon^2\right).
\end{align}

The deflection angle can be calculated by substituting the determinant~\eqref{The darkest evening of the year} and Gaussian curvature~\eqref{He gives his harness bells a shake} into Eq.~\eqref{DA-XY}, resulting in
\begin{align}
	\label{To ask if there is some mistake}
	\delta=&\frac{b_0}{b}\pm\frac{4J}{b^2v}+\mathcal{O}\left(\epsilon^2\right),
\end{align}
where, the positive and negative signs correspond to retrograde and prograde particle rays, respectively. The obtained result \eqref{To ask if there is some mistake} aligns with Eq. (A9) in Ref. \cite{massiveGB-CGJ}, derived using Werner's method in coordinates $(r,\phi)$. In comparison, the expressions for the Riemannian metric and its Gaussian curvature in the coordinate system $(r,\phi)$ presented in Ref. \cite{massiveGB-CGJ} are more complex, making calculations more cumbersome than those presented here.

Letting $v\to 1$ in Eq. \eqref{To ask if there is some mistake} yields the light deflection angle
\begin{align}
	\delta=&\frac{b_0}{b}\pm\frac{4J}{b^2}+\mathcal{O}\left(\epsilon^2\right).
\end{align}
This result was initially obtained by Jusufi and \"{O}vg\"{u}n~\cite{Fog lifted early I worked in the garden} using Werner's method in the coordinates $(r,\phi)$, and their computation was also cumbersome in comparison.

\endgroup

\section{Conclusion}
\label{Conclusion}

In this paper, we have introduced Cartesian-like coordinates $(X,Y)$, making Werner's method a convenient approach for calculating deflection angles in the weak-field approximation. Typically, the calculations of Werner's method in the coordinates $(r,\phi)$ are highly challenging, which has hindered its possible applications and obscured the advantages of the method itself. This work effectively eliminates this obstacle.

We have established a general procedure for implementing Werner’s method in Cartesian-like coordinates. As an application, we computed the deflection angle of massive particles in Kerr spacetime, rotating Bardeen (Hayward) 
regular spacetime, and Teo wormhole spacetime, respectively. Compared to the literature that utilized Werner’s method in \((r, \phi)\) coordinates for the same problems, we found that our calculations are much easier. In the Cartesian-like coordinates \((X, Y)\), each coordinate is treated equally, and the components of the osculating 
Riemannian metric in these coordinates have the same dimensions, which is not the case in the coordinates \((r, \phi)\). 
The equality of coordinates \(X\) and \(Y\) may be the reason why Werner’s method is convenient to use in Cartesian-like coordinates.

To emphasize our method, this paper only considers asymptotically Euclidean osculating Riemannian spaces and is limited to the leading-order deflection angle. For further research, we propose three directions. First, we aim to extend the Cartesian-like coordinate approach to the study of particle trajectory deflection in asymptotically non-flat spaces, such as rotating spacetimes with a deficit angle. In these cases, the deflection angle expression \eqref{Gibbons-Werner} requires modification~\cite{I knew no one worth my envying him,To think that once I was the same man did not embarrass me}, which consequently necessitates adjustments to Eq.~\eqref{DA-XY}. Second, a key challenge is computing higher-order deflection angles using Werner’s method. This remains particularly difficult in the coordinates $(r,\phi)$, and no corresponding work currently exists. However, the reduced computational complexity in Cartesian-like coordinates makes this feasible. Furthermore, a comparison between Eqs.~\eqref{DAF} and \eqref{KangjuHLbyT} reveals that Cartesian-like coordinates are better suited for calculating higher-order deflection angles. In  the coordinates $(X,Y)$, only low-order particle trajectories need to be considered, eliminating the need to iterate over low-order deflection angles and thereby simplifying the calculation process. 

Third, an important avenue for future research is the study of gravitational lensing in Finsler spaces beyond the Randers type. With increasing interest in Finsler gravity, it is likely that gravitational lensing in general Finsler spaces will become relevant. Due to the applicability of the osculating Riemannian framework, Werner’s method is expected to play a role in such scenarios. Moreover, we believe that the introduction of Cartesian-like coordinates will alleviate the computational challenges associated with this potential investigation, thereby promoting further advancements in this field.  


\end{document}